\documentclass[prl,a4paper,twocolumn,superscriptaddress,longbibliography]{revtex4-1}

\usepackage{amssymb}
\usepackage{amsmath}
\usepackage{amsfonts}
\usepackage{graphicx}
\usepackage{bm}
\usepackage{xcolor}
\usepackage{multirow}
\usepackage{comment}

\usepackage[final]{pdfpages}
\usepackage{pgffor}

\makeatletter
\AtBeginDocument{\let\LS@rot\@undefined}
\makeatother

\definecolor{darkblue}{rgb}{0.1,0.2,0.6}
\definecolor{darkred}{rgb}{0.8,0.1,0.2}
\definecolor{darkgreen}{rgb}{0,0.6,0.1}
\usepackage[colorlinks,citecolor=darkblue,linkcolor=darkred,urlcolor=darkblue]
{hyperref}
\usepackage[all]{hypcap}
\usepackage[normalem]{ulem}
\usepackage{mathtools}
\usepackage{amsmath}%
\usepackage{amsfonts}%
\usepackage{amssymb}
\newcommand{\bg}{ \begin{gather} }
\newcommand{\eg}{\end{gather}}
\newcommand{\be}{ \begin{equation} }
\newcommand{\ee}{\end{equation}}
\newcommand{\bea}{ \begin{eqnarray} }
\newcommand{\eea}{\end{eqnarray}}

\begin{document}

\title{Emergent continuous symmetry in anisotropic flexible two-dimensional materials}

\author{I. S. Burmistrov}

\affiliation{L. D. Landau Institute for Theoretical Physics, Semenova 1-a, 142432, Chernogolovka, Russia}
\affiliation{Laboratory for Condensed Matter Physics, National Research University Higher School of Economics, 101000 Moscow, Russia}

\author{V. Yu. Kachorovskii}

\affiliation{A. F.~Ioffe Physico-Technical Institute, 194021 St.~Petersburg, Russia}

\author{M. J. Klug} 

\affiliation{Institute for Theory of Condensed Matter, Karlsruhe Institute of Technology, 76131 Karlsruhe, Germany}

\author{J. Schmalian}

\affiliation{Institute for Theory of Condensed Matter, Karlsruhe Institute of Technology, 76131 Karlsruhe, Germany}

\affiliation{\mbox{Institute for Quantum Materials and Technologies,
Karlsruhe Institute of Technology, 76131 Karlsruhe, Germany}}

\begin{abstract}
We develop the theory of anomalous elasticity in two-dimensional flexible materials with orthorhombic crystal symmetry.    
Remarkably, in the universal region, where characteristic length scales are larger than the rather small Ginzburg scale  ${\sim} 10\, {\rm nm}$, these materials possess an infinite set of flat phases
which are connected by emergent continuous symmetry. This hidden symmetry leads to the formation of a stable line of fixed points corresponding to different phases. The same symmetry also enforces power law scaling with momentum of the anisotropic bending rigidity and Young's modulus,
controlled   by a single universal exponent -- the very same along the whole line of fixed points. These anisotropic flat phases are uniquely labeled by the ratio of absolute Poisson's ratios. We apply our theory to monolayer black phosphorus  (phosphorene).
\end{abstract}

\maketitle


The discovery of graphene~\cite{Novoselov2004,Novoselov2005,Zhang2005} and closely related atomically thick materials \cite{Novoselov2012} led to the  field of flexible two-dimensional (2D) materials \cite{2Dmat}.  The hexagonal crystal symmetry of graphene results in  elastic and electronic transport properties identical to those of isotropic system.
More recently, research has shifted towards other 2D materials, including 2D black phosphorus (phosphorene) \cite{Ling2015,Galluzzi2020}, transition metal dichalcogenide monolayers \cite{Wang2018,Durnev2018}, and metal monochalcogenide monolayers \cite{Sarkar2020,Barraza-Lopez2021}. Because of their different crystal structure, these 2D materials demonstrate anisotropic physical properties, such as electron and thermal transport, optical absorption,  photolumenescence, Raman scattering, and -- as will be important in this paper --  elastic response.
In particular, for  2D materials with  orthorhombic crystal symmetry, e.g. phosphorene, metal monochalcogenide monolayers (SiS, SiSe, GeS, GeSe, SnS, SnSe), monolayers GeAs$_2$, WTe$_2$, ZrTe$_5$, Ta$_2$NiS$_5$, etc.~\cite{Li2019},  the elastic free energy   {\it does not} reduce to that of an isotropic crystalline membrane and results in anisotropic Young's moduli and Poisson's ratios. 

 The idea of anomalous elasticity of isotropic crystalline membranes dates back to the seminal work by Nelson and Peliti \cite{Nelson1987}.    
Later it was found that the competition between anomalous elasticity and thermal fluctuations in clean membranes leads to the existence of a transition from  a flat to a
crumpled phase with increasing of  temperature \cite{Aronovitz1988,Paczuski1988,David1988,Aronovitz1989,Guitter1989,Doussal1992}. Currently there is substantial interest in furthering our  understanding of clean  crystalline membranes  \cite{Kats2014,Kats2016,Burmistrov2016,Kosmrlj2017,Burmistrov2018a,Burmistrov2018b,Saykin2020,Coquand2020,Mauri2020,Mauri2021}.

\begin{figure}[b]
\minipage{0.52\columnwidth}\includegraphics[width=\textwidth]{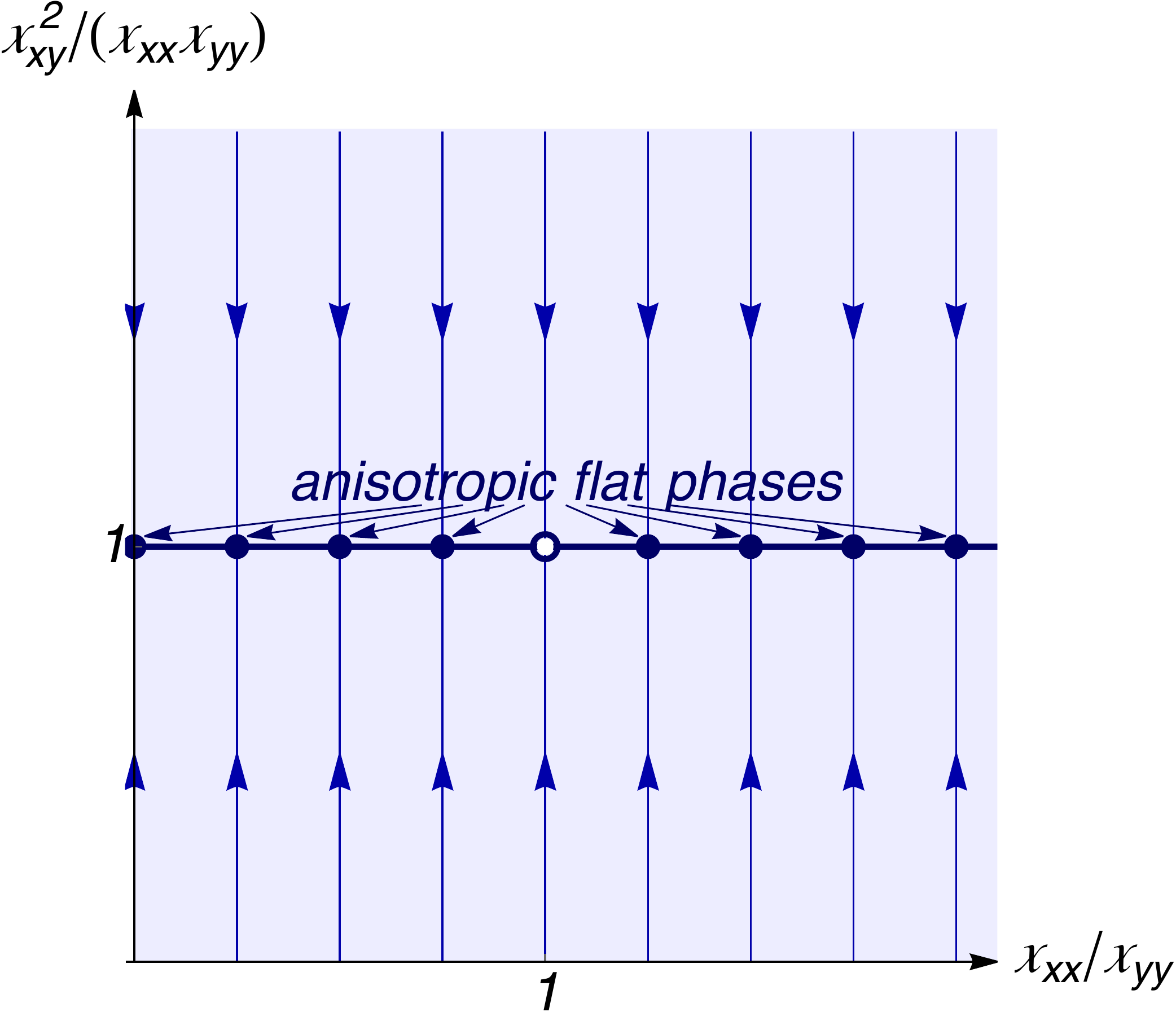}\endminipage
\quad
\minipage{0.43\columnwidth}\includegraphics[width=\textwidth]{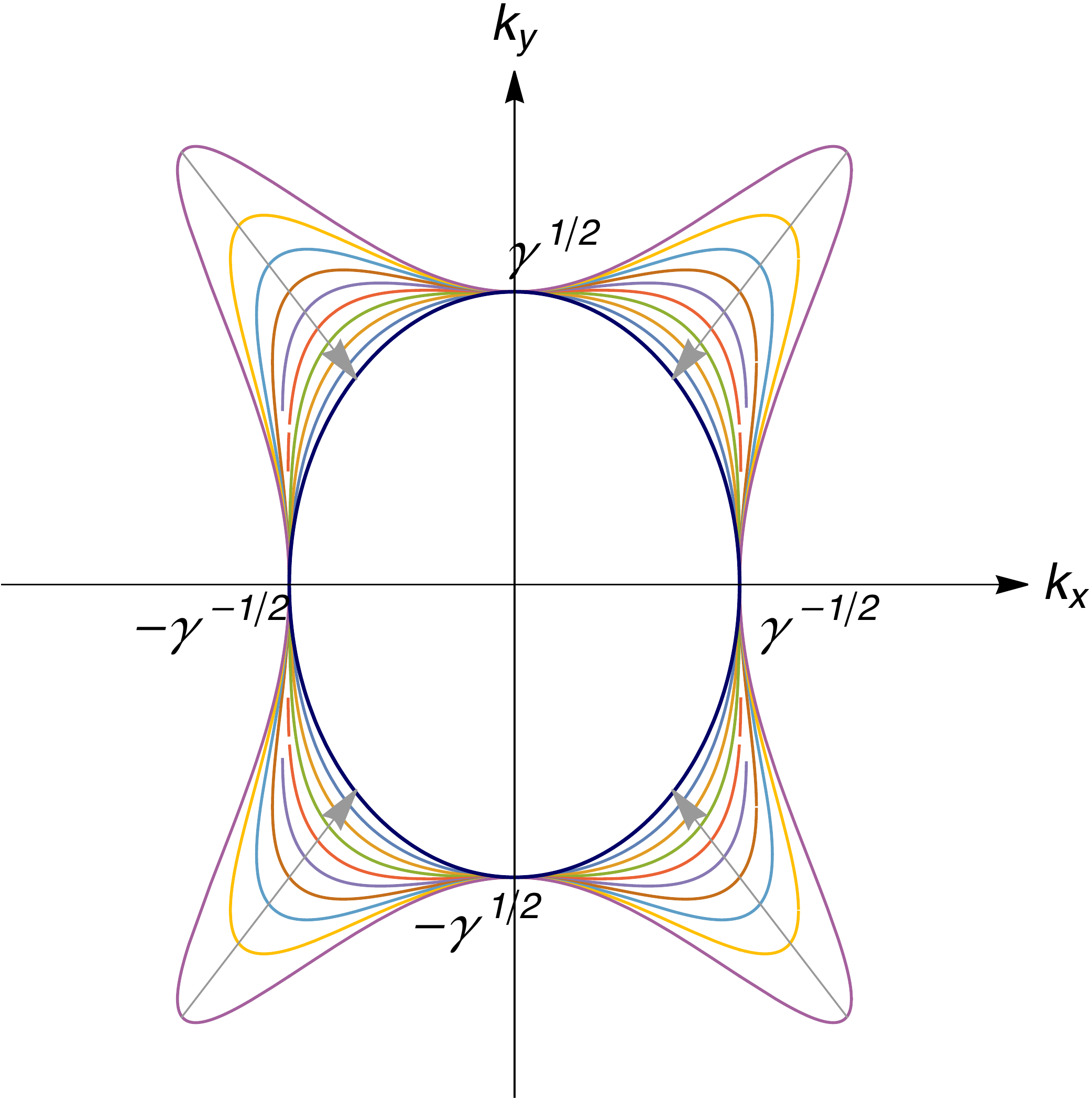}\endminipage
\caption{Left: Sketch of the RG flow (arrows indicate direction of increase of a system size $L$). Anisotropic flat phases on the line of fixed points are marked by blue dots. The isotropic flat phase is indicated by the white point. Right: 
The contour plot of schematic change of the normalized bending rigidity $k^4 \varkappa(\theta_{\bm{k}})/\sqrt{\varkappa_{xx}\varkappa_{yy}}$ under the RG flow for $\varkappa_{xx}{>}\varkappa_{yy}$, Eq. \eqref{eq:kappa:Y:eff}. For $\varkappa_{xx}{<}\varkappa_{yy}$ one needs to interchange the axes $k_x$ and $k_y$.}
\label{Fig:kappa}
\end{figure}

In this Letter we develop the theory of anomalous elasticity of 2D membranes with orthorhombic crystal symmetry. Our theory is focused on the universal regime when the typical size, $L$, of the membrane is large in comparison with the so-called Ginzburg scale, $q_*^{-1}$. As we show, for the mentioned 2D materials $q_*^{-1}{\sim}10\, {\rm nm}$ is extremely small, making the universal regime experimentally highly relevant. 
We predict that for $L{\gg} q_*^{-1}$, in contrast to graphene and dichalcogenide monolayers, these 2D materials 
possess  an {\it infinite set} of flat phases with
universal size-dependent anisotropic elastic properties. 
Although these phases perfectly preserve the orthorhombic anisotropy, they are connected by an emergent continuous symmetry. The latter ensures that the anisotropic anomalous Hooke's law is controlled by the same universal exponent for all phases.
Analyzing  the effective interaction of soft flexural phonons, these results are obtained from renormalization group (RG) equations that govern scaling of the bending rigidity tensor $\varkappa_{\alpha\beta}$ with $L$.
 Here $\alpha,\beta {\in} \{x,y\}$ are spatial indices. We obtain a {\it  line of stable fixed points} that describes an infinite set of  anisotropic flat phases corresponding to different degrees  of orthorhombicity $\varkappa_{xx}/\varkappa_{yy}$ (see Fig.~\ref{Fig:kappa}).  
 In all cases the angular-dependent 
 bending rigidity, $\varkappa(\theta_{\bm k}){=}\varkappa_{\alpha\beta}^{(0)} \hat{k}_\alpha^2\hat{k}_\beta^2$, where $\hat{\bm k}{=}\left(\cos\theta_{\bm{k}},\sin\theta_{\bm{k}}\right)$ is the unit vector along the momentum $\bm{k}$, flows to an elliptic  dependence on $\theta_{\bm k}$ controlled by the ratio $\gamma{=}(\varkappa_{xx}^{(0)}/\varkappa_{yy}^{(0)})^{1/4}$. The power law scaling with $k$ is  then governed by the universal exponent $\eta$ (see Fig. \ref{Fig:kappa} and Eqs.~\eqref{eq:kappa:Y:eff}). The emergent continuous symmetry ensures that $\eta$ is the critical exponent  for a flat isotropic membrane \cite{bookNelson}.      
Finally, we study the transition to a tubular phase (discussed earlier in Refs.\cite{Toner1989,Radzihovsky1995,Bowick1997,Radzihovsky1998,Essafi2011}), where an anisotropic 2D membrane is crumpled along one direction. We demonstrate that for a generic anisotropic membrane such transition occurs inevitably with increase of temperature.
However,    for  the  realistic above-mentioned 2D materials it happens at unphysically high temperature (of the order of tens of eV).

\noindent\textsc{\color{blue} Model.} --- The free energy describing thermal fluctuations in the flat phase of a  2D membrane 
with  orthorhombic crystal symmetry  can be written as \cite{Toner1989}
\begin{gather}
\mathcal{F} = \frac{1}{2}\int d^2 \bm{x} \Bigl [\varkappa_{\alpha\beta}^{(0)}  (\nabla_\alpha^2 \bm{r})(\nabla_\beta^2 \bm{r})
    + c_{11} u_{xx}^2 +
 c_{22} u_{yy}^2\notag \\
  + 2c_{12}u_{xx}u_{yy}
+ 4c_{66}u_{xy}^2 \Bigr ].
\label{eq:action:0}
\end{gather}
Here, $u_{\alpha\beta} {=} (\partial_\alpha \bm{r}\partial_\beta \bm{r}{-}\delta_{\alpha\beta})/2$ where $\bm{r}$ is a $d{=}d_c{+}2$ dimensional vector parametrizing  the membrane. For physical membranes, $d_c{=}1$ while anharmonic effects can be efficiently described in terms of an expansion in $1/d_c$. 
The parameters $\{c_{\alpha\beta},c_{66}\}$ denote the elastic moduli of 2D crystalline material. In the case of $\varkappa_{xx}^{(0)}{=}\varkappa^{(0)}_{yy}$ and $c_{11}{=}c_{22}$, the tetragonal crystal symmetry holds. For graphene which has the hexagonal symmetry, the bending energy is isotropic,  $\varkappa_{xx}^{(0)}{=}\varkappa_{yy}^{(0)}{=}\varkappa^{(0)}_{xy}$ together with $c_{11}{=}c_{22}{=} \lambda{+}2\mu$, $c_{12}{=}\lambda$, and $c_{66}{=}\mu$. For orthorhombic systems we allow for generic $c_{\alpha\beta}$ and $c_{66}$.

We choose the following 
parametrization of the coordinates: $r_1{=}\xi_x x + u_x$, $r_2{=}\xi_y y +u_y$, and $r_{a+2}{=}h_a$ with $a{=}1,\dots, d_c$, such that $u_{\alpha\beta}= (\xi_\alpha^2-1)\delta_{\alpha\beta}/2+\tilde{u}_{\alpha\beta}$, where
\begin{equation}
\tilde{u}_{\alpha\beta}= \frac{1}{2}\Bigl (
\xi_\beta\partial_\alpha u_\beta +\xi_\alpha\partial_\beta u_\alpha+\partial_\alpha \bm{h}\partial_\beta \bm{h} + \partial_\alpha \bm{u}\partial_\beta \bm{u} \Bigr ).
\label{eq:u:alpha:beta}
\end{equation}
Here  no summation over repeating indices is implied. The vectors $\bm{u}{=}\{u_x,u_y\}$ and $\bm{h}{=}\{h_1,\dots,h_{d_c}\}$ stand for in-plane and out-of-plane displacements, respectively. Assuming low 
enough temperatures (see below), we can neglect the term  $\partial_\alpha \bm{u}\partial_\beta \bm{u}$ in comparison with $\partial_\alpha \bm{h}\partial_\beta \bm{h}$ in Eq. \eqref{eq:u:alpha:beta}. Then, following Ref. \cite{Nelson1987}, we integrate over $\bm{u}$ and obtain the effective free energy written in terms of the out-of-plane phonons only \footnote{Here a `prime' sign in the last integral indicates that the interaction with $q{=}0$ is excluded.},
\begin{align}
\mathcal{F} = & \frac{1}{8}\int d^2\bm{x}\, c_{\alpha\beta} \varepsilon_\alpha\varepsilon_\beta  +  \frac{1}{2}\int \frac{d^2\bm{k}}{(2\pi)^2}
\varkappa_0(\theta_{\bm{k}}) k^4 \bm{h}_{\bm{k}}\bm{h}_{-\bm{k}}
 \notag \\ + & \frac{1}{8} \int \frac{d^2\bm{q}}{(2\pi)^2} Y_0(\theta_{\bm{q}}) 
 \Biggl | \int \frac{d^2\bm{k}}{(2\pi)^2} \frac{[\bm{k}\times\bm{q}]^2}{q^2} \bm{h}_{\bm{k+q}}\bm{h}_{-\bm{k}}\Biggr |^2 ,
\label{eq:action:2}
\end{align}
where we introduce
$\varepsilon_\alpha{=}\xi_\alpha^2{-}1{+} \int \! \frac{d^2\bm{k}}{(2\pi)^2}
k^2_\alpha \bm{h}_{\bm{k}}\bm{h}_{-\bm{k}}$.
We assume that the following  inequalities hold $\varkappa_{xx}^{(0)}, \varkappa_{yy}^{(0)}{>}0$ and  $\varkappa_{xy}^{(0)}{>}{-}(\varkappa_{xx}^{(0)} \varkappa_{yy}^{(0)})^{1/2}$. They guarantee that the bare angle--dependent bending rigidity $\varkappa_0(\theta){>}0$ for all angles $\theta$, such that the membrane is stable against transition into a tubular phase at zero temperature.  
The bare value of the Young modulus 
reads ($\epsilon_{\alpha\beta}$ is  fully antisymmetric tensor) \cite{Wei2014}
\begin{equation}
Y_0(\theta_{\bm q}) =c_{66}\Bigl [\hat{q}_x^2\hat{q}_y^2+\frac{c_{66}
c_{\alpha\beta}\epsilon_{\alpha\alpha^\prime}\epsilon_{\beta\beta^\prime}}{c_{11}c_{22}-c_{12}^2} \hat{q}_{\alpha^\prime}^2 \hat{q}_{\beta^\prime}^2 \Bigr ]^{-1} .
\end{equation}

\begin{figure}[t]
\centerline{\includegraphics[width=0.45\textwidth]{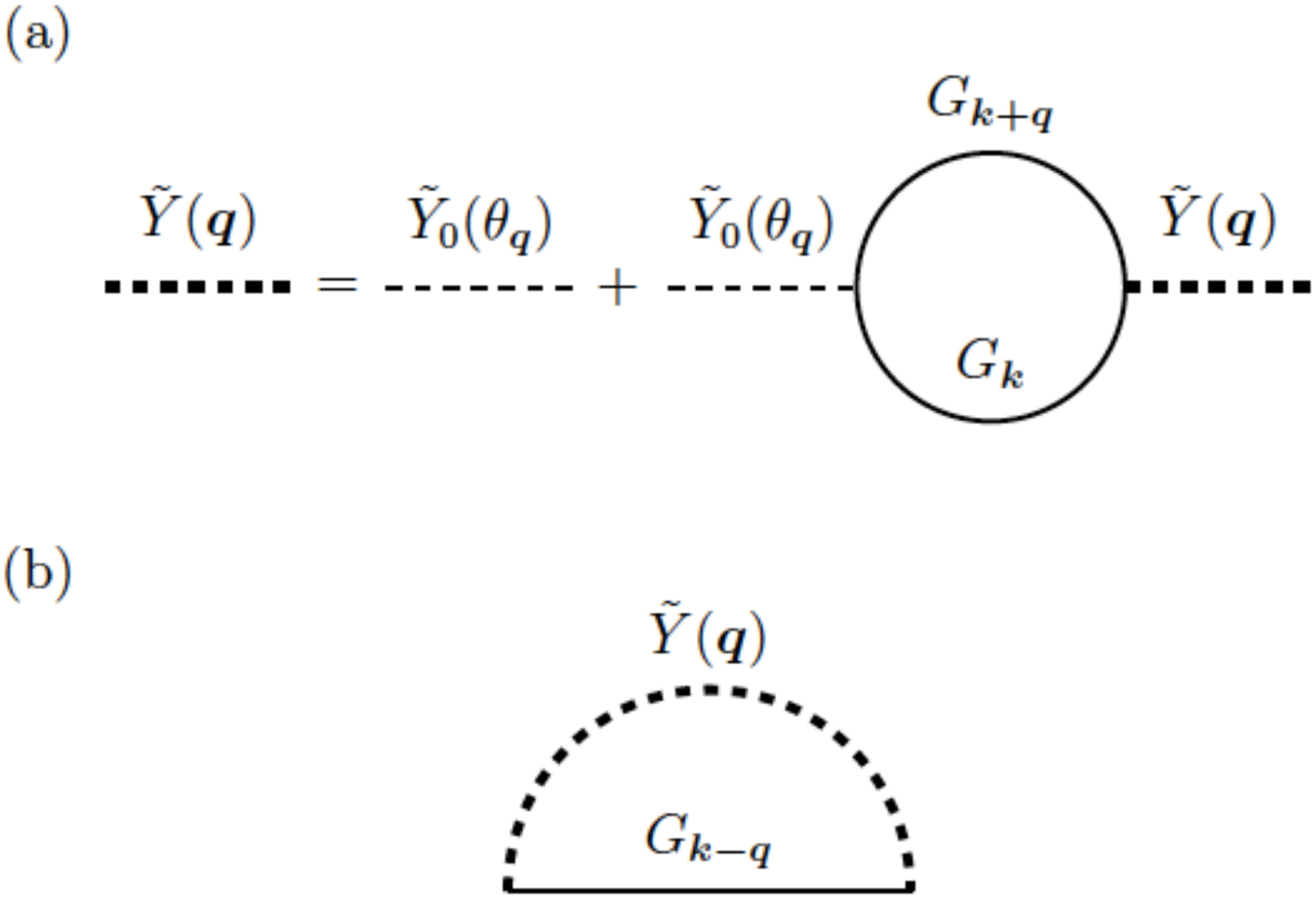}}
\caption{(a) The RPA-type resummation for the Young's modulus. (b)
The self-energy correction to first order in $1/d_c$. The solid line represents the bare Green's function $G_{\bm{k}}$. The thin (thick) dashed line denotes the (bare) screened interaction.}
\label{Figure-D1}
\end{figure}

The bending rigidity in Eq. \eqref{eq:action:2} is subject to renormalization, $\varkappa_0(\theta){\to} \varkappa(\theta)$, and, consequently, shows anomalous scaling.
This can be fully described by the following 
running coupling constants,
\begin{equation}
\gamma = \left(\!\frac{\varkappa_{xx}}{\varkappa_{yy}}\!\right)^{1/4}\!\!\!, \quad \tilde\varkappa{=}(\varkappa_{xx}\varkappa_{yy})^{1/2},\quad
 t = \frac{\tilde\varkappa-\varkappa_{xy}}{3\tilde\varkappa +\varkappa_{xy}} ,
 \end{equation}
 \color{black}
such that $0{<}\gamma{<}\infty$ and $|t| {<} 1$. 
The parameter $\gamma$ controls asymmetry between $x$ and $y$ axes existing  in the orthorhombic symmetry class. The case $\gamma{=}1$ corresponds to the tetragonal symmetry, where   $t_0$ describes the tetragonal distortion of the bending energy of the   membrane.  
By rescaling of  the momenta, 
\color{black}
\begin{equation}
k_x \mapsto  k_x/\sqrt{\gamma}, \qquad k_y\mapsto k_y \sqrt{\gamma}. 
\label{eq:kxky:transform}
\end{equation}
one can represent the bending energy in Eq. \eqref{eq:action:2} in a fashion formally similar to  a system with the tetragonal symmetry,
\begin{equation}
\varkappa(\theta) \mapsto \tilde{\varkappa}(\theta)=
\tilde{\varkappa}[1+t \cos(4\theta)]/(1+t) .
\label{eq:kappa:new}
\end{equation}
We emphasize that after the   rescaling \eqref{eq:kxky:transform} the free energy $\mathcal{F}$ has still a symmetry that is lower than the tetragonal one due to the quartic term in $h$. The modified Young's modulus $\tilde{Y}_0(\theta)$ 
still depends explicitly on $\gamma$. 
However, the elastic properties of membranes in the universal region  are independent of a particular form of  $\tilde{Y}_0(\theta)$\cite{Nelson1987,Aronovitz1988,Aronovitz1989}.

\noindent\textsc{\color{blue}Renormalization.} --- For $d_c{\gg} 1$ the effective free energy $\mathcal{F}$ can be analyzed by treating the quartic interaction perturbatively. The necessary information can be extracted from the exact two-point Green's function  $\langle h_i(\bm{k}) h_j(-\bm{k})\rangle {\equiv} \mathcal{G}_{\bm{k}}
\delta_{ij}$ where the average is taken with respect to the free energy $\mathcal{F}$.
The quadratic part of $\mathcal{F}$ determines the bare Green's function 
$G_{\bm{k}} {=} T/[\tilde{\varkappa}_0(\theta_{\bm k}) k^4]$. 

As usual, the bare interaction, $\tilde{Y}_0(\theta)$, between flexural phonons is screened by RPA-type diagrams (see Fig.~\ref{Figure-D1}a). The screened interaction becomes independent of $\tilde{Y}_0(\theta)$ in the long wave limit,  
$q{\ll}q_*$ where  $q_*^{-1}{\sim} [\varkappa^{(0)}_{xx}\varkappa^{(0)}_{yy}/(d_c \lambda T)]^{1/2}$,  is the  Ginzburg length.
Here $\lambda$ denotes a typical value of the elastic moduli. 
Therefore, at $q{\ll}q_*$ the theory \eqref{eq:action:2} 
becomes independent of the 
coupling constant $\gamma$ 
as a consequence of emergent  
hidden symmetry. 
The screened interaction behaves as $q^2/d_c$ at $q{\to} 0$. This facilitates construction of  the regular perturbation theory in $1/d_c$ for the self-energy $\Sigma_{\bm k} {=} G^{-1}_{\bm k}{-}\mathcal{G}^{-1}_{\bm k}$  (see diagram in Fig.~\ref{Figure-D1}b). 
The perturbation theory for $\Sigma_{\bm{k}}$ has infrared logarithmic divergences as $k{\to}0$. They can be used to extract the RG behavior of the bending rigidity $\tilde\varkappa(\theta)$(see Supplemental Material \cite{SM}). 

\begin{figure}[t]
\centerline{\includegraphics[width=0.38\textwidth]{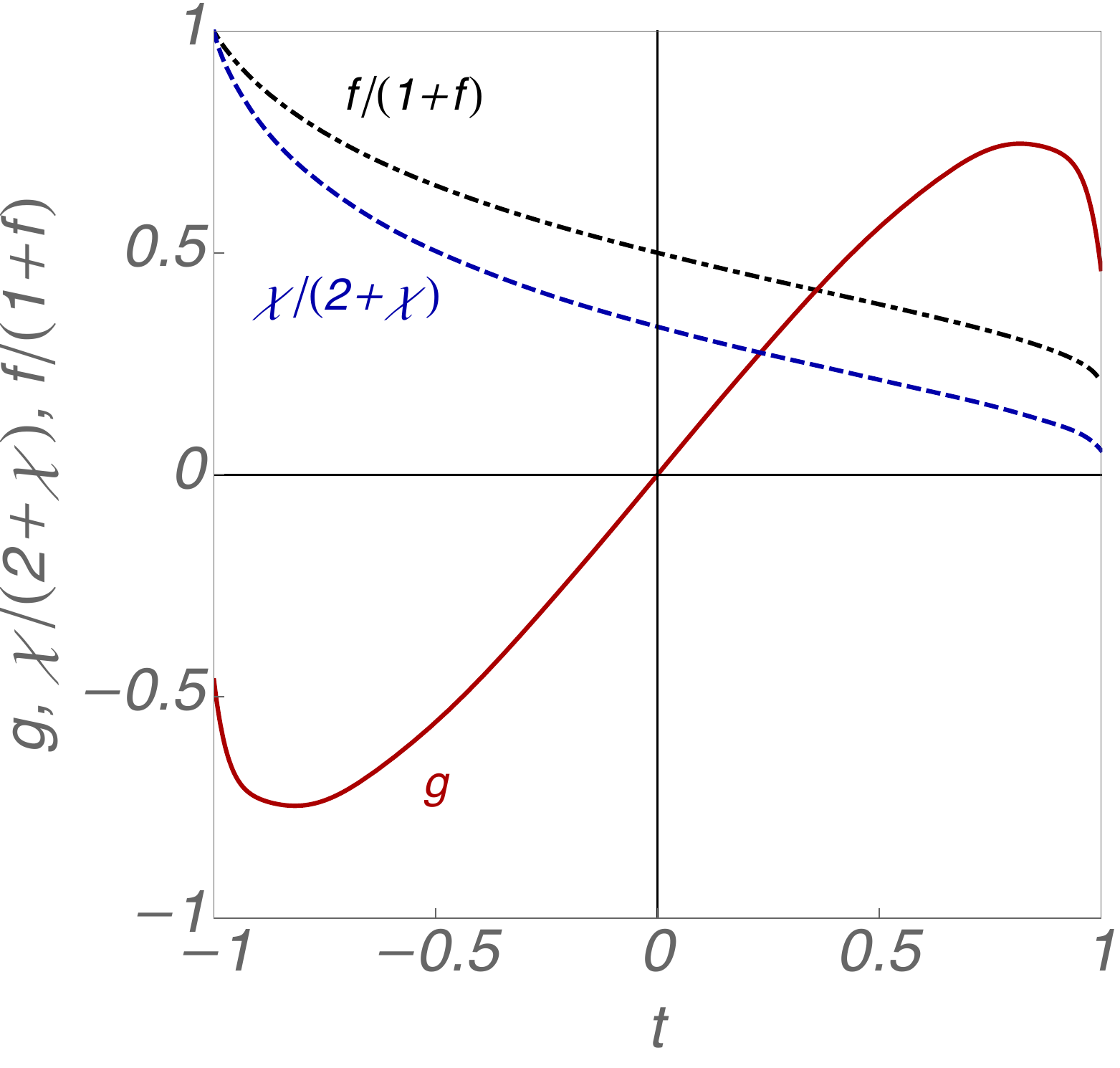}}
\caption{The functions $g(t)$ and $\chi(t)$, that enter the RG Eqs. \eqref{eq:RG:eqs}, and the function $f(t)$ that determines transition temperature to the tubular phase are shown.
}
\label{Figure-D2}
\end{figure}

The RG equations (to the lowest order in $1/d_c$) can be written in the following form ($\Lambda{=}\ln (q_*/k)$),
\begin{equation}
\frac{d\gamma}{d\Lambda}=0, \quad \frac{dt}{d\Lambda} \simeq - \frac{2}{d_c} g(t), \quad \frac{d\ln \tilde\varkappa}{d\Lambda}
\simeq \frac{2}{d_c} \chi(t) ,
\label{eq:RG:eqs}
\end{equation}
where $t(0){=}t_0$ and $\tilde\varkappa(0){=}\tilde\varkappa_0$. We stress that the first of these equations is a consequence of the hidden symmetry and is not limited by $1/d_c$ expansion. We note also that $g(t)$ is an odd function. 
RG functions $g$ and $\chi$  tend to constants as $t{\to} 1$ and
 have the following asymptotic expansions at $|t|{\ll}1$:
$g(t){\simeq}(65 t/54){+}O(t^3)$ and $\chi(t){\simeq}  
1{-}(65 t/54){+}(145 t^2/162){+}O(t^3)$ (see Fig.~\ref{Figure-D2}). 
For $t{\to}{-}1$ the function $g(t)$ tends to the constant whereas $\chi(t){\sim}1/(1{+}t)$. 
There exists an infra-red stable line of fixed points at $t{=}0$ at which the bending rigidity scales just like for isotropic membranes,  $\tilde{\varkappa}{\sim}(q_*/k)^\eta$, where $\eta{\simeq} 2/d_c$.  Eqs. \eqref{eq:RG:eqs} implies that the line of fixed points can only be reached asymptotically, $t{\sim}(k/q_*)^\psi$, with exponent $\psi {\approx} 65/(27 d_c)$. This crossover exponent determines the rate at which the bending rigidity approaches the elliptical form; see Fig.~\ref{Fig:kappa}.

\noindent\textsc{\color{blue} Beyond $1/d_c$ expansion.} --- The stable line of fixed points at $t{=}0$ emerges as the RG equations for $t$ and $\tilde\varkappa$ in Eqs.\eqref{eq:RG:eqs} are independent of $\gamma$ (see Fig. \ref{Fig:kappa}). We reiterate that the existence of the line of the fixed points is a direct consequence of the emergent hidden symmetry of the theory in the universal regime, $q{\ll}q_*$, and, thus, goes beyond the lowest order expansion in $1/d_c$.  Hence, such line of fixed points exists for physical membranes with $d_c{=}1$. For arbitrary $d_c$ (including $d_c{=}1$) it is natural to assume that (i) the line of stable fixed points remains stable ($\psi{>}0$), (ii) there are no higher harmonics in $\tilde{\varkappa}(\theta)$ are generated under the RG flow, (iii) $t{=}0$ is the only fixed point in the interval $0{\leqslant} t{<}1$. Then, the elasticity of 2D membranes with the orthorhombic crystal symmetry in the universal regime can be deduced from anomalous elasticity of isotropic membranes.

Making inverse rescaling to Eq.~\eqref{eq:kxky:transform}, we find  that in the original coordinate system the anomalous elasticity is  described by the  effective bending rigidity and Young's modulus:
\begin{equation}
\begin{split}
\varkappa_{\rm eff}(\bm{k}) & \sim \Bigl(\gamma \cos^2 \theta_{\bm{k}}+\gamma^{-1} \sin^2 \theta_{\bm{k}}\Bigr)^{2-\eta/2} \left (\frac{q_*}{k}\right )^{\eta} , \\
Y_{\rm eff}(\bm{q}) & \sim \left[ \Bigl(\gamma \cos^2 \theta_{\bm{q}}+\gamma^{-1} \sin^2 \theta_{\bm{q}}\Bigr)^{-1} \frac{q^2}{q^2_*}\right ]^{1-\eta}  .
\end{split}
\label{eq:kappa:Y:eff}
\end{equation} 
At first glance, these equations obey  only the  discrete $\mathbb{Z}_2$ symmetry  of  orthorhombic  phase, which is simply the discrete symmetry of an ellipse. Remarkably,   there  is  also   a hidden continuous symmetry related to the above mentioned rescaling.  Indeed,   after  rescaling, the anisotropic  problem with    $\varkappa_{xx}$ and $\varkappa_{yy}$ is reduced to the isotropic one. 
The hidden symmetry is just an arbitrary rotation of this effective isotropic model.    

The scaling of 
$\varkappa_{\rm eff}$ and $Y_{\rm eff}$  with the absolute value of momentum is controlled by the critical exponent $\eta$, 
known for the isotropic membrane. For $d_c{\gg} 1$ it is given as $\eta{\simeq} 2/d_c{+} (73-68 \zeta(3))/(27 d_c^2){+}O(1/d_c^3)$ \cite{Saykin2020}, whereas for $d_c{=}1$ the numerics predicts $\eta{=}0.795\pm 0.01$ \cite{Troster2013}.
 The change of the angular dependence of the bending rigidity upon the RG flow is illustrated in Fig. \ref{Fig:kappa}. 
For $\gamma{=}1$, $\varkappa_{\rm eff}$ and $Y_{\rm eff}$ become independent of the angle. Thus a membrane with the tetragonal crystal symmetry is equivalent to an isotropic membrane in the universal regime. 

The mechanics of a membrane on the line of the fixed points remains anisotropic. Under application of an unidirectional tension, $\sigma_{x}$, along the $x$ axis the following deformations along the $x$ and $y$ axes are induced, 
\begin{equation}
\delta\xi_x^2 \sim \gamma^{-1}(\sigma_x/\gamma)^{\alpha}, \qquad  \delta\xi_y^2 \sim \gamma(\sigma_x/\gamma)^{\alpha} .
\label{eq:xi:x:y}
\end{equation}
Here the critical exponent $\alpha$, which controls anomalous Hooke's law \eqref{eq:xi:x:y}, is expressed via $\eta$ exactly in the same way as in an isotropic membrane, $\alpha{=}\eta/(2-\eta)$. As expected, the membrane is deformed easier in the direction for which the bending rigidity is smaller. For example, from Eq. \eqref{eq:xi:x:y} it follows that $\delta\xi_y^2{>}\delta\xi_x^2$ for
$\gamma{>}1$ ($\varkappa_{xx}^{(0)}{>} \varkappa_{yy}^{(0)}$).
We note that the power-law behavior \eqref{eq:xi:x:y} holds for $\sigma_x {\ll} \sigma_* {\sim}  \tilde \varkappa_0 q_*^2$.
The results for unidirectional tension along the $y$ axis can be obtain from Eq. \eqref{eq:xi:x:y} under the following interchange $x {\leftrightarrow} y$ and $\gamma{\leftrightarrow} 1/\gamma$. 

The anomalous Hooke's law \eqref{eq:xi:x:y} results in anisotropic {\it negative} absolute Poisson's ratio. In the anisotropic flat phase one finds for $\sigma_x, \sigma_y {\ll}\sigma_*$,
\begin{equation}
\nu_x = - \frac{\delta\xi_y^2(\sigma_x)}{\delta\xi_x^2(\sigma_x)} = \gamma^2 \nu, \quad 
\nu_y = - \frac{\delta\xi_x^2(\sigma_y)}{\delta\xi_y^2(\sigma_y)} = \frac{\nu}{\gamma^2} .
\label{eq:PR:n}
\end{equation}
Here $\nu {\approx} {-}1 {+}2/d_c {-}a/d_c^2 {+}O(1/d_c^3)$ with $a{\approx} 1.76\pm 0.02$ denotes the absolute Poisson's ratio for the isotropic membrane \cite{Saykin2020}. We mention that the following relation holds, 
$\nu_x/\nu_y {=}\gamma^4$. Therefore, the measurement of the absolute Poisson's ratios in the regime of anomalous Hooke's law, $\sigma_x, \sigma_y {\ll}\sigma_*$, allows one to uniquely characterize the anisotropy in bending rigidity.

\noindent\textsc{\color{blue}Tubular phase.} --- Usually, the crumpling transition is deduced from the equation of states that relates the stretching factor and the external tension. In the absence of the latter the equation of states, $\partial \langle \mathcal{F}\rangle /\partial \xi_\alpha^2{=}\langle\varepsilon_\alpha\rangle{=}0$,  yields the dependence of $\xi_\alpha^2$ on temperature. Instead, following Ref.  \cite{Gornyi:2015a}, we introduce the momentum-dependent stretching factor, $\xi_\alpha^2(k){=}1{-}d_c \int\!\frac{d^2{\bm q}}{(2\pi)^2}\Theta(q{-}k)q_\alpha^2 \mathcal{G}_{\bm q}$, 
that includes contributions from the flexural phonons with momenta $q$ larger than a given momentum $k$. The dependence $\xi_x^2(k)$ can be cast in the form of the RG equation,
\begin{equation}
\frac{d\xi^2_x}{d\Lambda} = \frac{d_c T}{4\pi \varkappa_{xx}^{3/4}\varkappa_{yy}^{1/4}}\left (\frac{1+t}{1-t}\right )^{1/2} , \quad \xi_x^2(\Lambda=0)=1 .
\label{eq:xi:RG}
\end{equation}
The RG equation for $\xi^2_y$ can be obtained from Eq. \eqref{eq:xi:RG} by interchange $x$ and  $y$. 
At low temperatures the flat phase, in which $\xi_\alpha^2(k)$ is positive for all $k{<}q_*$, is realized. For $\gamma{<}1$,  with increase of temperature $\xi_x^2(k)$ vanishes at some finite value of $k$. At the same time $\xi_y^2(k)$ is still positive for all $k{<}q_*$. Therefore, the tubular phase exists above the transition temperature $T_{\rm x} {=}
T_{\rm x}^{(0)} f(t_0)$ where $T_{\rm x}^{(0)}{=}(8\pi /d_c^2)[\varkappa_{xx}^{(0)3}\varkappa_{yy}^{(0)}]^{1/4}$  
stands for the temperature of the crumpling transition at $t_0{=}0$. 
We note that  
the crumpling occurs along the direction, $x$, that corresponds to the 
smaller bare bending rigidity, $\varkappa_{xx}^{(0)}{<} \varkappa_{yy}^{(0)}$.
The function $f(t)$ can be found from RG equations \eqref{eq:RG:eqs} and \eqref{eq:xi:RG} \cite{SM}. We note that  $f(t)$ is a monotonously decreasing function with $f(t{\to}{-}1){\sim}(1+t)^{-1}$, $f(0){=}1$, and $f(1){\approx} 0.3$.  The function $f(t)$ is shown in Fig. \ref{Figure-D2}.  With even more increase of temperature the tubular phase experienced crumpled transition \cite{Radzihovsky1995,Radzihovsky1998}.
For $\gamma{<}1$ the tubular phase at $T{>}T_{\rm y}$ corresponds to  $\xi^2_y(k{=}0){=}0$. The transition temperature $T_{\rm y}$ can be obtained from the expression for $T_{\rm x}$ upon interchange $x$ and $y$. 

\noindent\textsc{\color{blue}Discussion.} --- As illustration of our theory 
we employ it to 2D black phosphorus. We are not aware of direct measurements of elastic and bending moduli of phosphorene. Recent numerical calculations reports the following magnitudes: 
$c_{11}{\approx}105.2$, $c_{22}{\approx}26.2$, $c_{12}{\approx}18.4$, and $c_{66}{\approx}22.4$ (measured in N/m) \cite{Wang2015} as well as  $\varkappa_{xx}^{(0)}{\approx}8.0$ eV and $\varkappa_{yy}^{(0)}{\approx}4.8$ eV \cite{Zhang2015-BP}.
This yields the inverse Ginzburg length $q_*{\approx} 0.1$ nm$^{-1}$. Hence,  anomalous elasticity should 
dominate elastic properties of available experimentally micron size samples. We remind that the Young's modulus and bending rigidity of graphene are $340$ H/m and $1.4$ eV, respectively, such that the inverse Ginzburg length at the room temperature is of the order $1$ nm$^{-1}$. The anisotropy parameter can be estimated as $\gamma{\approx}1.1$, indicating that the effective bending rigidity and Young's modulus, Eqs. \eqref{eq:kappa:Y:eff}, are expected to be slightly anisotropic.

For 2D black phosphorus our theory predicts that nonlinear Hooke's law, Eq. \eqref{eq:xi:x:y}, and negative absolute Poisson's ratios, Eq. \eqref{eq:PR:n}, should be observable for tensions smaller than $\sigma_*{\approx} 10^{-2}$ N/m. Although we are not aware of measurements of strain--stress dependence of phosphorene we believe that it can be performed in a way similar to graphene \cite{Nicholl2015}. 
There are several computations of the Poisson's ratios for black phosphorus from  first principles \cite{Jiang2014,Du2016}. While these results yield  a negative Poisson's ratios of phosphorene, the numerical computation of the Poisson's ratio in the regime of low applied tensions, $\sigma_{x,y}{\ll}\sigma_*$, 
may suffer from a problem with proper boundary conditions in a finite size samples \cite{Burmistrov2018b}. As in the case of graphene, a direct measurement of Poisson's ratio of phosphorene is challenging.
We estimate the transition temperature to the tubular phase, $T_{\rm x}^{(0)}$,
to be of the order of $50$ eV making anisotropic flat phase of 
phosphorene to be absolutely stable  
from the experimental point of view. 

We also note that results \eqref{eq:kappa:Y:eff} for the effective  bending rigidity and Young's modulus are important for an electron transport in 2D materials. Recently, the anisotropy of the carrier mobility of 2D black phosphorus was studied using an effective  bending rigidity of precisely the asymptotic form of Eq. \eqref{eq:kappa:Y:eff} \cite{Rudenko2016,Brener2017}.

It is worthwhile to mention that recently, 
distinct anisotropic model which breaks the  $O(d)$ rotational invariance in the embedded space has been studied \cite{Doussal2021}. 
It would be interesting to study the effect of crystalline anisotropy discussed above in the model of Ref. \cite{Doussal2021}.

To summarize, we developed the theory of anomalous elasticity in systems with orthorhombic crystal symmetry, relevant for a large number of recently studied 2D flexible materials. Our key finding is the hidden symmetry of the theory emerging in the universal regime that leads to existence  of the infinite number of anisotropic flat phases in the long-wave limit. 
These flat phases have anisotropic bending rigidity and Young's modulus whereas the scaling with momentum is the same as in the isotropic case. 
They are uniquely labeled by the ratio of absolute Poisson's ratios in the two perpendicular directions.  Our theory can easily be extended to even less symmetric 2D materials with triclinic and monoclinic crystal symmetries, e.g. monolayers ReS$_2$, ReSe$_2$, GaTe, GeP, GeAs, SiP, SiAs, etc. Also, one can extended our theory to include the effects of in-plane and curvature disorder. The later was shown to be important for qualitative explanation of nonlinear strain--stress relation in graphene \cite{Gornyi2016}.

\begin{acknowledgements}
We thank M. Glazov and V. Lebedev for useful comments and discussions. The work was funded in part by the Alexander von Humboldt Foundation, by the Russian Ministry of Science and Higher Educations, the Basic Research Program of HSE,  by the Russian Foundation for Basic Research, grant No. 20-52-12019, and by the Deutsche Forschungsgemeinschaft (DFG, German Research Foundation) project SCHM 1031/12-1.
\end{acknowledgements}

\bibliography{biblio}



\section*{Supplementary material}

# Emergent continuous symmetry in anisotropic flexible two-dimensional materials

I. S. Burmistrov,[1,2] V. Yu. Kachorovskii,[3] M. Klug,[4] and J. Schmalian[4,5]

[1] L. D. Landau Institute for Theoretical Physics, Semenova 1-a, 142432, Chernogolovka, Russia
[2] Laboratory for Condensed Matter Physics, National Research University Higher School of Economics, 101000 Moscow, Russia
[3] A. F. Ioffe Physico-Technical Institute, 194021 St. Petersburg, Russia
[4] Institute for Theory of Condensed Matter, Karlsruhe Institute of Technology, 76131 Karlsruhe, Germany
[5] Institute for Quantum Materials and Technologies, Karlsruhe Institute of Technology, 76131 Karlsruhe, Germany

In this notes, we present (i) a derivation of the renormalization group equations to the lowest order in $1/d_c$ (Eq. 8 of the main text) and (ii) a derivation of the flat–to–tubule transition temperature.

## S.I. RENORMALIZATION GROUP EQUATIONS TO THE LOWEST ORDER IN $1/d_c$

### A. Effective interaction

Within RPA-type resummation scheme (see Fig. 2 of the main text) we obtain the screened interaction as

$$\tilde{Y}(\boldsymbol{q}) = \frac{\tilde{Y}_0(\theta_{\boldsymbol{q}})/2}{1 + 3\tilde{Y}_0(\theta_{\boldsymbol{q}})\Pi_{\boldsymbol{q}}/2}, \tag{S1}$$

where the bare Young's modulus depends on the parameter $\gamma$,

$$\tilde{Y}_0(\theta_{\boldsymbol{q}}) = 4\lambda_{xyxy}\left[\sin^2(2\theta) + 4\lambda_{xyxy}\frac{(\gamma^{-2}\lambda_{xxxx}\cos^4\theta - (\lambda_{xxyy}/2)\sin^2(2\theta) + \lambda_{yyyy}\gamma^2\sin^4\theta)}{\lambda_{xxxx}\lambda_{yyyy} - \lambda_{xxyy}^2}\right]^{-1} \tag{S2}$$

Here the polarization operator is given as

$$\Pi_{\boldsymbol{q}} = \frac{d_c}{3T}\int\frac{d^2\boldsymbol{k}}{(2\pi)^2}\frac{[\boldsymbol{k}\times\boldsymbol{q}]^4}{q^4}G_{\boldsymbol{k}+\boldsymbol{q}}G_{-\boldsymbol{k}} = \frac{d_c T}{q^2}\mathcal{P}(\theta_{\boldsymbol{q}}), \quad \mathcal{P}(\theta_{\boldsymbol{q}}) = \frac{1}{3}\int\frac{d^2\boldsymbol{k}}{(2\pi)^2}\frac{[\boldsymbol{k}\times\boldsymbol{q}]^4}{q^2}\frac{1}{(\boldsymbol{k}+\boldsymbol{q})^4\tilde{\varkappa}_0(\theta_{\boldsymbol{k}+\boldsymbol{q}})k^4\tilde{\varkappa}_0(\theta_{-\boldsymbol{k}})}. \tag{S3}$$

Such form of the screened interaction is justified to the lowest order in $1/d_c$. Similar to the isotropic case, in the long wave vector limit, $q \ll q_*$, the effective interaction of flexural phonons becomes

$$\tilde{Y}(\boldsymbol{q}) \to 1/(3\Pi_{\boldsymbol{q}}), \tag{S4}$$

We note that in the absence of the anisotropy, $t_0=0$, the polarization operator is given as $\mathcal{P}(\theta_{\boldsymbol{q}})=1/(16\pi\tilde{\varkappa}_0^2)$.

### B. Renormalization group equation

To the lowest order in $1/d_c$ the self-energy is given by diagram shown in Fig. 2 of the main text

$$\Sigma_{\boldsymbol{k}} = -\frac{2}{3}T\int\frac{d^2\boldsymbol{q}}{(2\pi)^2}\frac{[\boldsymbol{k}\times\boldsymbol{q}]^4}{q^4}\frac{G_{\boldsymbol{k}-\boldsymbol{q}}}{\Pi_{\boldsymbol{q}}} = -\frac{2}{3d_c}\int\frac{d^2\boldsymbol{q}}{(2\pi)^2}\frac{[\boldsymbol{k}\times\boldsymbol{q}]^4}{q^2}\frac{1}{|\boldsymbol{k}-\boldsymbol{q}|^4\tilde{\varkappa}_0(\theta_{\boldsymbol{k}-\boldsymbol{q}})\mathcal{P}(\theta_{\boldsymbol{q}})}. \tag{S5}$$

The expression (S5) is logarithmically divergent. Within logarithmic accuracy at $k \ll q_*$, we obtain

$$\Sigma_{\boldsymbol{k}} \simeq -\frac{1}{3\pi d_c}\left(k^4\ln\frac{q_*}{k}\right)\int\limits_0^{2\pi}\frac{d\theta_{\boldsymbol{q}}}{2\pi}\frac{\sin^4(\theta_{\boldsymbol{k}}-\theta_{\boldsymbol{q}})}{\tilde{\varkappa}_0(\theta_{\boldsymbol{k}-\boldsymbol{q}})\mathcal{P}(\theta_{\boldsymbol{q}})} = -\frac{1}{24\pi d_c}\left(k^4\ln\frac{q_*}{k}\right)\int\limits_0^{2\pi}\frac{d\theta_{\boldsymbol{q}}}{2\pi}\frac{1}{\tilde{\varkappa}_0(\pi-\theta_{\boldsymbol{q}})\mathcal{P}(\theta_{\boldsymbol{q}})}$$

$$\times\left[3 - 4\cos(2\theta_{\boldsymbol{k}})\cos(2\theta_{\boldsymbol{q}}) + \cos(4\theta_{\boldsymbol{k}})\cos(4\theta_{\boldsymbol{q}})\right]. \tag{S6}$$

Here we used that $\theta_{\boldsymbol{k}-\boldsymbol{q}} \to \pi - \theta_{\boldsymbol{q}}$ as $k \to 0$.

It is convenient to introduce the dimensionless functions ($m=0,1,2$)

$$F_{2m}(t_0) = \frac{1}{16\pi} \int_0^{2\pi} \frac{d\theta}{2\pi} \frac{(1+t_0)\cos(2m\theta)}{\tilde{\varkappa}_0 \tilde{\varkappa}_0(\pi-\theta)\mathcal{P}(\theta)}. \tag{S7}$$

We note that the normalization is such that $F_0(t_0=0)=1$ and $F_2(t_0=0)=F_4(t_0=0)=0$. Moreover, the symmetry of $\tilde{\varkappa}_0(\theta)$ implies that the function $F_2$ is zero identically, $F_2(t_0)\equiv 0$ whereas $F_0(t_0)$ ($F_4(t_0)$) is the even (odd) function of $t_0$.

Converting perturbative result (S6) into the renormalization group equations, we find (see Eq. (8) in the main text)

$$\begin{aligned}\frac{dt}{d\Lambda} &= -\frac{2}{d_c}g(t), \qquad g(t) = tF_0(t) - F_4(t)/3, \\ \frac{d\ln\tilde{\varkappa}}{d\Lambda} &= \frac{2}{d_c}\chi(t), \qquad \chi(t) = \frac{F_0(t)+F_4(t)/3}{1+t},\end{aligned} \tag{S8}$$

where $\Lambda = \ln(q_*/k)$.

### C. Asymptotic expressions for $F_0$ and $F_4$ for $0\leqslant t<1$

Performing integration in Eqs. (S3) and (S7), one finds the following asymptotic expressions at $t\ll 1$,

$$\mathcal{P}(\theta) \simeq \frac{(1+t)^2}{16\pi\tilde{\varkappa}_0^2}\left[1 + \frac{2}{9}\cos(4\theta)t + \frac{85-2\cos(8\theta)}{90}t^2 + \frac{\cos(4\theta)(86+3\cos(8\theta))}{315}t^3 + O(t^4)\right] \tag{S9}$$

and

$$F_0(t) \simeq 1 - \frac{25}{81}t^2 + O(t^4), \quad F_4(t) \simeq -\frac{11}{18}t + \frac{527}{9720}t^3 + O(t^5). \tag{S10}$$

Using the above results one can derive the asymptotic results for the functions $g(t)$ and $\chi(t)$ mentioned in the main text.

For $1-t \ll 1$, the integral over the angle $\theta_{\boldsymbol{k}}$ in Eq. (S3) is dominated by vicinity of $\pi/4$ and $5\pi/4$. At the same time the integral over absolute value of momentum is dominated by $k \sim q\sin(\theta_q+\pi/4)$. Then, evaluating the integral over $\boldsymbol{k}$ in Eq. (S3), we find

$$\pi(\theta) \simeq \frac{1}{24}\frac{|\cos(2\theta)|}{1-t}, \qquad |\theta - \pi/4| \gg \sqrt{1-t}. \tag{S11}$$

In the limit $|\theta-\pi/4|\ll\sqrt{1-t}$, one can obtain that $\pi(\theta)\sim 1/\sqrt{1-t}$.

For $1-t\ll 1$ the integrals over $\theta$ in definitions of functions $F_0$ and $F_4$, Eq. (S8), are dominated by the region $\theta \simeq \pi/4$. Hence, we find ($\theta = \pi/4 + \delta\theta$)

$$F_0(t) \simeq -F_4(t) \simeq \frac{24}{32\pi^2}\int_{\sqrt{1-t}}^\infty \frac{d(\delta\theta)(1-t)}{(1-t+8\delta\theta^2)|\delta\theta|} \sim \int_{\sim 1}^\infty \frac{d(\delta\theta)}{(1+8\delta\theta^2)|\delta\theta|} \tag{S12}$$

Therefore, the functions $F_0(t)$ and $F_4(t)$ tend to a constant as $t\to 1$.

### S.II. TRANSITION TO THE TUBULAR PHASE

Using (8) and (12) in the main text, we find

$$\xi_x^2(\Lambda) = 1 - \frac{T}{T_{\text{x}}^{(0)}}\int_t^{t_0} \frac{d\tau}{g(\tau)}\left(\frac{1+\tau}{1-\tau}\right)^{1/2}\exp\left(-\int_\tau^{t_0} du\frac{\chi(u)}{g(u)}\right) \tag{S13}$$

where $T_\text{x}^{(0)} = (8\pi/d_c^2)[\varkappa_{xx}^{(0)3} \varkappa_{yy}^{(0)}]^{1/4}$. We note that $\Lambda$ enters the right hand side of the above equation via $t \equiv t(\Lambda)$. For $T > T_\text{x}$, where

$$T_\text{x} = T_\text{x}^{(0)} f(t_0), \qquad f(t_0) = \left[ \int_0^{t_0} \frac{d\tau}{g(\tau)} \left( \frac{1+\tau}{1-\tau} \right)^{1/2} \exp\left( -\int_\tau^{t_0} du \frac{\chi(u)}{g(u)} \right) \right]^{-1} \tag{S14}$$

the stretching factor $\xi_x^2$ vanishes at some finite value of $\Lambda$.

As one can check, the function $f(t_0)$ is monotonously decreasing function on the interval $-1 < t_0 < 1$. Using asymptotic results (S10), we find at $|t_0| \ll 1$,

$$f(t_0) \simeq 1 - t_0 + O(t_0^2). \tag{S15}$$

For $t_0 \to 1$ the function $f(t_0)$ tends to a constant, whereas for $t_0 \to -1$ we obtain $f(t_0) \sim 1/(1+t_0)$.



\end{document}